\documentclass[twocolumn,twoside,slac_two]{revtex4}
\usepackage{graphicx}
\usepackage{fancyhdr}
\begin{document}

\title{(Pseudo-) Scalar Operators in the MSSM and $B\to\phi K^*,
K\eta^{(\prime)}$ Decays}


\author{Hisaki~Hatanaka and Kwei-Chou~Yang}
\affiliation{Department of Physics, 
Chung-Yuan Christian University, Chungli, Taiwan}

\begin{abstract}
We study the effect of $b \to s \bar{s} s$ scalar/pseudoscalar
operators, originating from penguin diagrams of neutral Higgs bosons in
the MSSM, in $B \to K \eta^{(\prime)}$ and $B\to \phi K^*$ decays. These
operators can be Fierz-transformed into tensor operators, and the
resultant tensor operators could affect the transverse polarization
amplitudes in $B \to \phi K^*$ decays. We find that only when the weak
annihilations in $B \to \phi K^*$ into account, the polarization puzzle
can be resolved, so that new physics effects are strongly suppressed and
no more relevant to the enhancement of the transverse modes in $B \to
\phi K^*$ decays.  (This presentation is based on the work
\cite{Hatanaka-Yang})
\end{abstract}

\maketitle

\thispagestyle{fancy}

\newcommand{\dfrac}[2]{{\displaystyle\frac{#1}{#2}}}
\newcommand{\Ktwo}{K_2^*(1430)}
\section{Introduction}
In the $B \to VV$ ($V$ denotes a vector meson) decays, polarizations of
the two vector mesons have been extensively studied by experiments. The
decay amplitude for the B meson to two vector mesons can be
decomposed into three parts. In the transversity basis they are
longitudinal $A_0$ and two transverse modes, where the latter consist of
the parallel $A_\parallel$ and the perpendicular $A_\perp$. In terms of
them we can define the polarization fractions as $f_L = |A_0|^2$,
$f_\parallel =|A_\parallel|^2$ and $f_\perp = |A_\perp|^2$.  Here we
take the normalization: $|A_0|^2 +|A_\parallel|^2 + |A_\perp|^2 = 1$.
The naive factorization estimation yields $f_L : f_\parallel
: f_\perp = 1-{\cal O}(\Lambda^2/m_b^2): {\cal O}(\Lambda^2/m_b^2) :
{\cal O}(\Lambda^2/m_b^2)$. However, in the penguin dominated $B\to VV$
decays it is not the case. For example, the polarization fractions in
$B\to \phi K^*(892)$ decays are
\cite{Aubert:2006uk,Chen:2005zv,CDF,HFAG}
\begin{eqnarray}
f_L(B^+ \to \phi K^{*+}) &=& 0.50\pm0.05,
\nonumber \\
f_L(B^0 \to \phi K^{*0}) &=& 0.484\pm0.034,
\nonumber \\
f_{\perp}(B^+ \to \phi K^{*+}) &=& 0.20\pm0.05,
\nonumber \\
f_{\perp}(B^0 \to \phi K^{*0}) &=& 0.256\pm0.032.
\end{eqnarray}
These results are very different from the naive expectations. Such
discrepancy between the theory and experiments in the $B\to VV$ decays
is referred as ``the polarization puzzle (anomaly)''.  One possibility
of resolving the puzzle is to resort to the new physics (NP). In
\cite{Das-Yang} it is pointed out that tensor operators may play an
essential role of helicity flipping of the quarks.  One of the
resolutions within the standard model (SM) framework is to take into
account the large weak-annihilation effect
\cite{Kagan,Yang05,Beneke2006}.  In \cite{Kagan,Yang05}, the author
pointed out that the magnitude of annihilation correction is order of
${\cal O}(1/m_b^2 \log^2 m_b/\Lambda_h)$, and the effect can interfere
with the longitudinal and transverse modes destructively and
constructively, respectively.  It is noted, however, that the
perturbative QCD (pQCD) yields $f_L \gtrsim 0.75$ even with annihilation
effects \cite{Li:2004ti}.  In this presentation, we take into account $B
\to \phi K^*$ and $B \to K \eta^{(\prime)}$ data to further constrain
the possible NP contributions.

%%%%%%%%%%%%%%%%%%%%%%%%%%%%%%%%%%%%%%%%%%%%%%%%%%%%%%%%%%%%%%%%%%%%%%%
\section{Formulation and Analysis}
In \cite{Das-Yang}, $b\to s\bar{s}s$ NP operators $O_{i}$
($i=11,\cdots,26$) are introduced to resolve the polarization puzzle.
These operators are not independent and can be related with each other
through the following Fierz transformations:
\begin{eqnarray}
&& O_{19} = -\frac{1}{2} O_{14}, \quad
O_{20} = -\frac{1}{2} O_{13}, \nonumber \\
&& O_{21} = -\frac{1}{2} O_{6},  \quad
O_{22} = -\frac{1}{2} O_{5},            \\
&& O_{23} = -4 O_{15} -8 O_{16}, \quad
O_{24} = -8 O_{15} -4 O_{16}, \nonumber \\
&& O_{25} = -4 O_{17} -8 O_{18}, \quad
O_{26} = -8 O_{17} -4 O_{18}. \nonumber
\end{eqnarray}

In the naive factorization, the $B\to VV$ decay amplitudes are sensitive
to tensor operators but insensitive to scalar ones. On the other hand,
for $B \to PP$ (where $P$ is a pseudo-scalar meson) decay amplitudes it
is the other way around.

\begin{figure}[tbp]
\caption{The ratio $(B-A)/B$ for various $\tan\beta$ and
$m_A$}\label{abratio}
\includegraphics[width=0.8\linewidth]{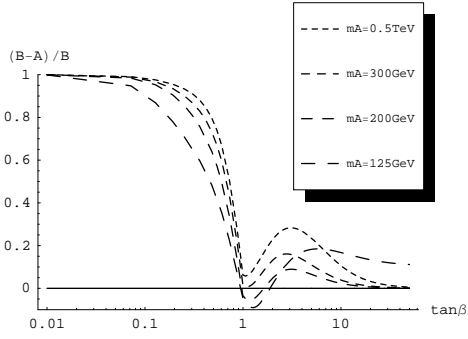}
\end{figure}

In the MSSM, scalar operators are induced by the neutral-Higgs penguin
diagrams \cite{Hatanaka-Yang,Huang}:
\begin{eqnarray}
c_{15} = D(A-B)\xi, \quad
c_{17} = D(A-B)\xi' ,
\nonumber\\
c_{19} = D(A+B)\xi, \quad
c_{21} = D(A+B)\xi' ,
\end{eqnarray}
where
\begin{eqnarray}
D &\equiv& \dfrac{1}{12\pi^2}
\dfrac{1}{\lambda_t}
\dfrac{e^2 g_s^2}{g^2 \sin^2\theta_W}
 f_b'(m_{\tilde{q}}^2 / m_{\tilde{g}}^2 ) m_s m_{\tilde{g}},
\nonumber \\
A &\equiv&
\dfrac{1}{m_{H^0}^2}
\left(\dfrac{\cos^2\alpha+(m_{H^0}^2/m_{h^0}^2)\sin^2\alpha
}{
\cos^2\beta}\right),
\nonumber \\
B &\equiv& \dfrac{1}{m_{A^0}^2}
\left(\dfrac{m_{A^0}^2}{m_{Z^0}^2} + \tan^2\beta \right),
\nonumber \\
\xi  &\equiv& \delta_{23}^{dLL}\delta_{33}^{dLR},
\quad
\xi' \equiv \delta_{23}^{dRR}\delta_{33}^{dLR*}.
\label{MSSM-param}
\end{eqnarray}
We find that $A$ and $B$ are not independent and constrained by
$-0.1 \lesssim (B-A)/B \le 1$ (see FIG.~\ref{abratio}).

Tensor operators are not directly induced in the MSSM and can be
obtained from scalar operators through the Fierz-transformations.  In
the $B \to K\eta^{(\prime)}$ decays, the NP effects modify the Wilson
coefficients $c_{5,6} \to c_{5,6} + \Delta c_{5,6}$:
\begin{eqnarray}
\Delta c_6 &=& \left\{
\begin{array}{cl} 
D B (\xi - \xi'), & \mbox{for } \alpha_4, \beta_3 \\
\frac{1}{2} [2-(B-A)/2], & \mbox{for } \alpha_3, \beta_2, \beta_{S3},
\end{array}\right.
\nonumber \\
\Delta c_5 &=& 0.
\end{eqnarray}
Here the $\alpha_i,\beta_j$ are defined in
\cite{Hatanaka-Yang,Beneke:2002jn,Beneke:2003zv}. In $B \to \phi K^{*}$
decays the NP effects modify the Wilson coefficients $c_{i} \to
\bar{c}_i$, where
\begin{eqnarray}
\bar{c}_6 - c_6 = 
\frac{1}{2} \left(\frac{B-A}{B} -2\right) DB\xi' ,
\nonumber \\
\bar{c}_{14} = 
\frac{1}{2} \left(\frac{B-A}{B} -2\right) DB\xi ,
\nonumber \\
c_{23} = \frac{1}{12} D(A-B)\xi,\quad
c_{24} = -\frac{1}{6} D(A-B)\xi, \nonumber \\
c_{25} = \frac{1}{12} D(A-B)\xi', \quad
c_{24} = -\frac{1}{6} D(A-B)\xi'.
\end{eqnarray}
The contributions of the NP tensor operators to the decay amplitudes in
 the transversity basis are given by
\begin{eqnarray}
\overline{A}_{0}^{NP} &=& -4 i f_\phi^T m_B^2 
[\tilde{a}_{23} - \tilde{a}_{25}]
[h_2 T_2(m_\phi^2)-h_3 T_3(m_\phi^2)],
\nonumber \\
\overline{A}_{\parallel}^{NP} &=& 4 i \sqrt{2} f_\phi^T m_B^2 
[\tilde{a}_{23} - \tilde{a}_{25}]
f_2 T_2(m_\phi^2),
\nonumber \\
\overline{A}_{\perp}^{NP} &=& 4 i \sqrt{2} f_\phi^T m_B^2 
[\tilde{a}_{23} + \tilde{a}_{25}]
f_1 T_1(m_\phi^2),
\end{eqnarray}
where
\begin{eqnarray}
\tilde{a}_{23} \equiv \bar{c}_{23} + \frac{1}{2} \bar{c}_{24} 
 + {\cal O}(\alpha_s) \simeq \frac{1}{8N_c} \frac{B-A}{B} DB \xi,
\nonumber \\
\tilde{a}_{25} \equiv \bar{c}_{25} + \frac{1}{2}\bar{c}_{25}
 + {\cal O}(\alpha_s) \simeq \frac{1}{8N_c} \frac{B-A}{B} DB \xi'.
\end{eqnarray}

In the $\chi^2$ fit, we have used 20 observables from $B^{0}\to\phi
K^{*0}$ and $B^{+}\to\phi K^{*+}$ data, and 7 observables from $B\to
K\eta^{(\prime)}$. As for NP fitting parameters we use $(B-A)/B$,
$DB\xi$, and $DB\xi'$.  For simplicity we consider the two NP scenarios:
(A) NP-scenario where $\xi\neq0$, $\xi'=0$ and (B) NP scenario where
$\xi=0$, $\xi'\neq0$.  We include the annihilation contributions in the
decay amplitudes \cite{Beneke2006}.
%
%%%%%%%%%%%%%%%%%%%%%%%%%%%%%%%%%%%%%%%%%%%%%%%%%%%%%%%%%%%%%%%%%%%%%%
\section{Results and Comments}

\begin{figure}[tbp]
\caption{ Contour plots for $\Delta \chi^2 \equiv \chi^2 - \chi^2_{min}
$ \cite{Hatanaka-Yang}.  Allowed regions of $\Delta \chi^2 \le 1$, $1
\le \Delta \chi^2 \le 4$ and $4 \le \Delta \chi^2 \le 9$ are shown by
dark, medium-dark and light-gray regions.  ``$\times$'' indicates the
location of the global minimum. The origin corresponds to the SM, and
the circle at the origin indicates the allowed upper-limit from the $B_s
\to \mu^+\mu^-$ data.  }\label{contour}
\includegraphics[width=0.8\linewidth]{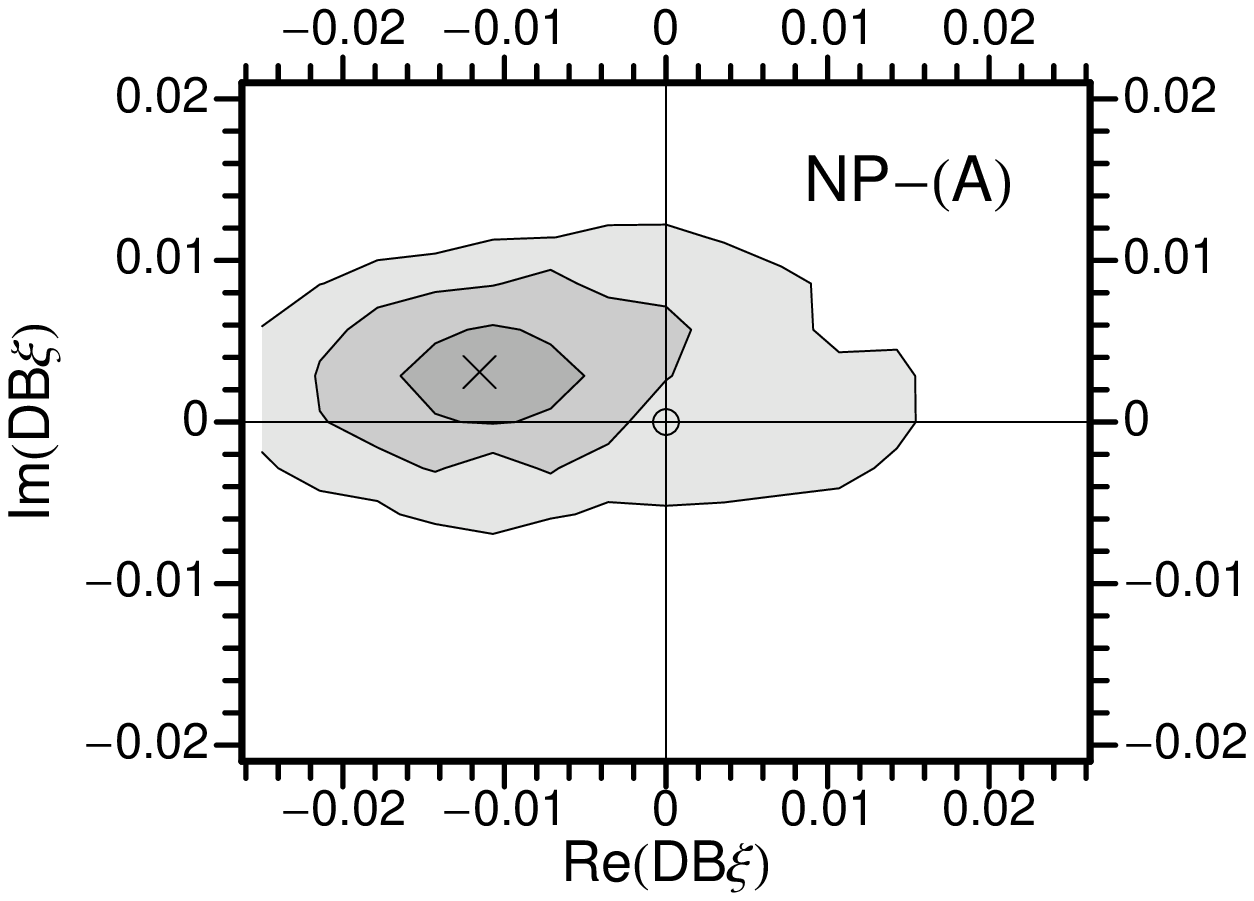}
\includegraphics[width=0.8\linewidth]{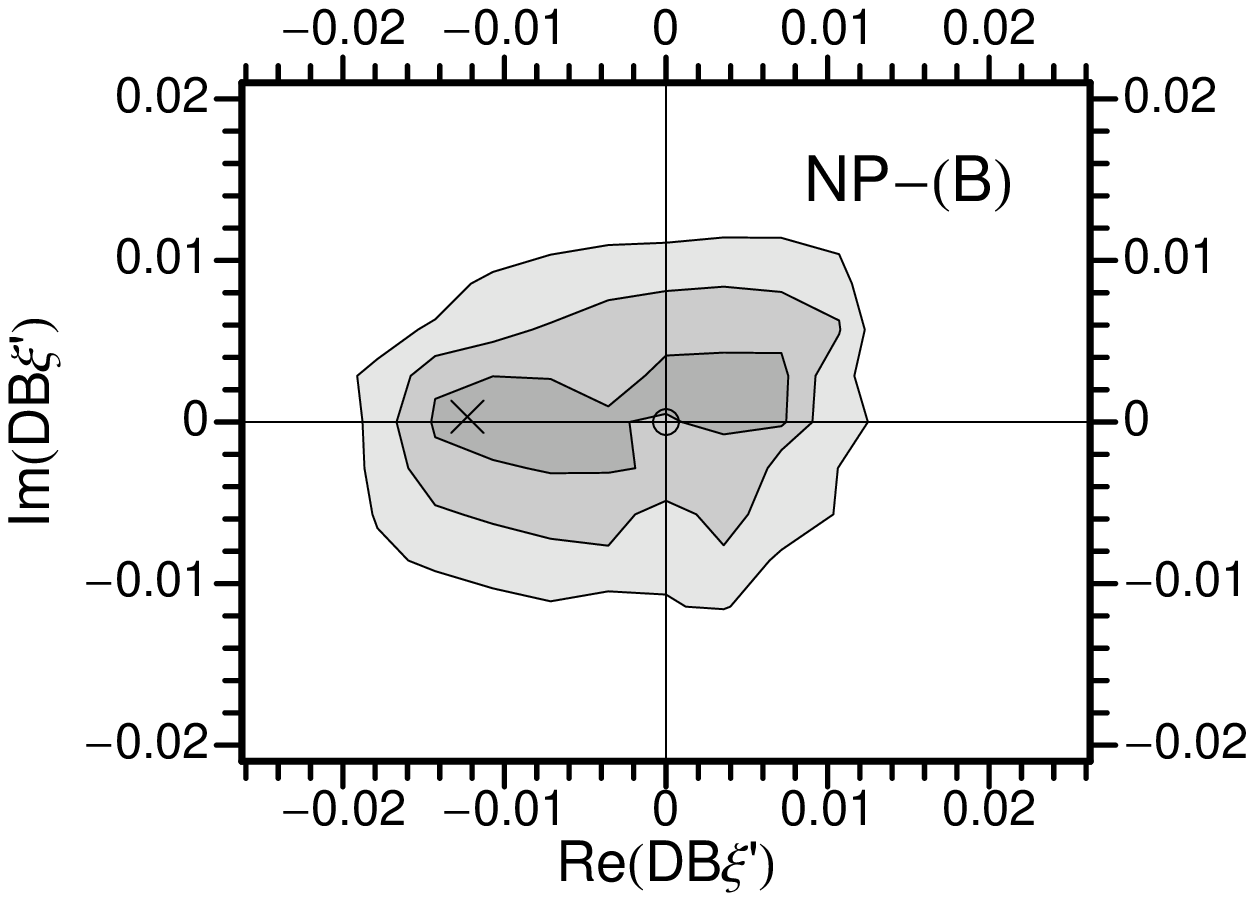}
\end{figure}

We take the annihilation effects into account and perform $\chi^2$
fitting. The results are $\chi^2_{\min}/\mbox{d.o.f} = 9.8/17$ for the
scenario (A) and $15.5/17$ for (B). The contour-plots of the $\chi^2$
fits are shown in FIG.~\ref{contour}.
It should be noted that, if ignoring the $B\to\phi K^*$ annihilation
contributions we could not find any solutions which explain both the
$B\to K\eta^{(\prime)}$ and $B\to\phi K^*$ data cannot be satisfied
simultaneously.

From the results for $DB\xi^{(\prime)}$ and the fact that $-0.1 \lesssim
 (B-A)/B \le 1$ we obtain
\begin{eqnarray}
|\tilde{a}_{23}| \le 7.1\times 10^{-4}, \quad
|\tilde{a}_{25}| \le 6.1\times 10^{-4},
\end{eqnarray}
which are much smaller than the values given in \cite{Das-Yang},
\begin{eqnarray}
|\tilde{a}_{23,\rm DY}| &=& 4.4^{+0.3}_{-0.2} \times 10^{-3}, 
\nonumber\\
|\tilde{a}_{25,\rm DY}| &=& 5.4^{+0.5}_{-0.3} \times 10^{-3},
\end{eqnarray}
 where the annihilation contributions are ignored.  We conclude that the
polarization puzzle can be resolved by the SM weak-annihilation effect
if the NP tensor operators are induced by the NP scalar ones.
Moreover, scalar operators can be strongly constrained by the
upper-bound of the $B_s \to \mu^+\mu^-$ decay \cite{bsmumu}:
\begin{eqnarray}
{\cal B}(B_s \to\mu^+\mu^-) \le 7.5\times10^{-8}.
\end{eqnarray}
In FIG.~\ref{contour} the bound is indicated as the small circle at the
origin.  Since in both scenarios the $B_s\to\mu^+\mu^-$ data and the SM
are located within contours where $\chi^2/\mbox{d.o.f.}$ is sufficiently
small, we thus conclude that the new-physics effects due to the scalar
operators may be negligible.

Finally we make a remark on the recently observed large polarization
fraction $f_L$ for $B \to \phi \Ktwo$ \cite{Aubert:2006uk}. If tensor
operators play an significant role in $B \to VT$ ($T$ denotes a tensor
meson) decays, $f_L$ may significantly deviate from unity. The current
experiment is consistent with our result since in our analysis the
tensor operator are found to be tiny.  However, in the present study we
cannot exclude the possibility that sizable NP effects contribute
directly to tensor operators, instead of scalar/pseudoscalar operators,
and, moreover, a cancellation may take place between weak annihilation
and contributions due to NP tensor operators in the $B\to \phi\Ktwo$
decay. For the point of view of the new physics, $B\to \phi\Ktwo$ may be
sensitive to the $B\to \Ktwo$ tensor form factor which can be further
explored from the $B \to \Ktwo\gamma$ decay.

\begin{acknowledgments}
This work is partly supported by 
National Science Council (NSC) of Republic of China
under Grant
NSC 96-2811-M-033-004 and
NSC 96-2112-M-033-004-MY3.

\end{acknowledgments}

\bigskip % extra skip inserted
%\begin{thebibliography}{9}   % Use for  1-9  references

\end{document}